\documentclass[12pt]{article}
\usepackage{times}
\usepackage{geometry}
\usepackage[utf8]{inputenc}
\usepackage{enumitem,amssymb}
\usepackage{ragged2e}
\newlist{thematic}{itemize}{8}
\setlist[thematic]{label=$\square$}
\usepackage{pifont}

\usepackage{url}
\usepackage{soul}
\usepackage{amssymb, amsmath, epsfig}
\usepackage{hyperref}
\usepackage{tabularx}
\usepackage{subfigure}
\usepackage{epstopdf}
\usepackage{comment}
\usepackage{floatrow}
\usepackage{framed}
\newfloatcommand{capbtabbox}{table}[][\FBwidth]
\usepackage[svgnames]{xcolor}
\usepackage{blindtext}
\usepackage{graphicx}
\usepackage{sidecap}
\usepackage{times}
\usepackage{mathptmx}
\usepackage{wrapfig}
\usepackage{titlesec}
\usepackage[yyyymmdd]{datetime}
\usepackage{hyperref}
\usepackage{ifthen}
\usepackage{afterpage}
\usepackage{appendix}
\usepackage{lineno}
\usepackage{multirow}
\usepackage{bm}
\usepackage{array}
\newcolumntype{L}[1]{>{\raggedright\let\newline\\\arraybackslash\hspace{0pt}}m{#1}}
\newcolumntype{C}[1]{>{\centering\let\newline\\\arraybackslash\hspace{0pt}}m{#1}}
\newcolumntype{R}[1]{>{\raggedleft\let\newline\\\arraybackslash\hspace{0pt}}m{#1}}
\usepackage{framed}
\definecolor{shadecolor}{named}{CornflowerBlue}

\usepackage[most]{tcolorbox}

\DeclareGraphicsExtensions{.ps}

\begin{document}

\raggedright
\huge
Astro2020 White Paper \linebreak
Theoretical Astrophysics 2020-2030
\linebreak
\normalsize

\noindent \textbf{Thematic Areas:}  \hspace*{10pt} $\square$ Ground-Based Project \hspace*{10pt} $\square$ Space-Based Project \hspace*{10pt}\\
$\boxtimes$ Infrastructure Activity \hspace*{10pt} $\boxtimes$ Technological Development Activity \\
$\boxtimes$ State of the Profession Consideration\linebreak

\textbf{Principal Author:} \\ 
Name: Juna A. Kollmeier\\
Institution: Carnegie Observatories, 813 Santa Barbara Street, Pasadena, CA 91101, USA\\
Email: jak@carnegiescience.edu
 \linebreak

\textbf{Co-authors:}\\
 Lauren Anderson, Flatiron Institute, landerson@flatironinstitute.org, Andrew Benson, Carnegie Observatories, abenson@carnegiescience.edu, Tamara Bogdanovi\'c, Georgia Tech, tamarab@gatech.edu, Michael Boylan-Kolchin, UT Austin, mbk@astro.as.utexas.edu, James S. Bullock, UC Irvine, bullock@uci.edu, Romeel Dav\'e, University of Edinburgh, rad@roe.ac.uk, Federico Fraschetti, University of Arizona, frasche@lpl.arizona.edu, Jim Fuller, Caltech, fuller@tapir.caltech.edu, Phil Hopkins, Caltech, ph@tapir.caltech.edu, Manoj Kaplinghat, UC Irvine, mkapling@uci.edu, Kaitlin Kratter, University of Arizona, kkratter@email.arizona.edu, Astrid Lamberts, Observatoire de la Côte d'Azur in Nice, M. Coleman Miller, University of Maryland, miller@astro.umd.edu, James E. Owen, Imperial College London, james.owen@imperial.ac.uk,  E. Sterl Phinney, Caltech, esp@tapir.caltech.edu, Anthony L. Piro, Carnegie Observatories, piro@carnegiescience.edu, Hans-Walter Rix, Max Planck Institute for Astronomy, rix@mpia.de, Brant Robertson, UC Santa Cruz, brant@ucsc.edu, Andrew Wetzel, UC Davis, awetzel@ucdavis.edu, Coral Wheeler, Caltech/Carnegie, coral@caltech.edu,  Andrew N. Youdin, University of Arizona, youdin@email.arizona.edu, Matias Zaldarriaga, Institute for Advanced Study, matiasz@ias.edu

\newpage

\noindent

\newpage

\justify
\section{Key Issue and Overview of Impact on the Field} 
\subsection{Issue}
The past two decades have seen a tremendous investment in observational facilities that promise to reveal new and unprecedented discoveries about the universe.  These machines are indeed marvels of technology and engineering and their large pricetags (all nearing or in excess of \$1B USD) are testament to the difficulty and expense of creating this equipment.  However, over the same period, no similarly large investment in theoretical work has taken place {\it even when accounting for computational infrastructure.}   

In this white paper, we argue that in order for the promised critical breakthroughs in astrophysics over the next decade and well beyond, the national agencies must take a serious approach to investment in theoretical astrophysics research.  We provide a multi-level strategy, from the grassroots to the national, to address the current under investment in theory relative to observational work.

\subsection{Impact on the Field}

\centerline{\it But happily, hypotheses, even if full of error, fail us not.  --- F.W. Argelander }
\bigskip
Theory plays several key roles within astronomy and astrophysics.  First and foremost, theory guides insight into the physical nature of the universe {\bf even when the theory is wrong}.  Indeed, it is the {\it difference} between theory and observation that has ushered in the most dramatic revolutions in physics and astronomy.  Examples abound ranging from the deviation of mercury's orbit relative to Newtonian dynamics that ultimately yielded General Relativity, to the variation in the expected cosmic microwave background radiation and the observed very smooth cosmic microwave background (CMB) that ultimately gave way to our current concordance model for the cosmos.

In the 20th century, it was perhaps sufficient to provide a small number of exceptional astrophysicists access to pencils and paper.  This approach no longer suffices in the 21st century.  Theoretical research has become increasingly complex and the community seeks to test these ideas at increasingly challenging observational regimes.  Many theorists perform heavily computational work, which, thus far, has been encouraged primarily in an ad-hoc, grass-roots level in individual departments and universities.  

Testing theoretical work is often used to motivate observing proposals and in justification for large missions.  Nevertheless, the support for this work is woefully lacking compared to the observations themselves.  Who is to make these predictions and who is to vet them when they increasingly rely on calculations that take months to years on large computers to reproduce?

Rather than continuing to give lip service to the need for theorists in the community, we call on the National Academy to take a strong stand in support of theory.  We call on NAS not only to acknowledge the critical role theory plays in leading the pathway to fundamental breakthroughs in dark matter and dark energy, but to endorse a specific program to support theory and thus keep astronomy a vibrant field of truth seeking, rather than a backdrop for technological research and development.

In addition to the obvious role of theory in making the nature of the universe manifest in clear, predictable, and manipulatable form, theoretical research serves two more practical functions.  The first is in the {\it Guidance of Observational Programs} and the second is in the {\it Interpretation of New Results}.  We discuss each of these in turn below.

\subsection{Guidance for Observations}

Some of the most successful observational programs of all time,  particularly in recent decades, were guided by theory.  These include the CMB, large scale structure of galaxies, gravitational lensing, gravitational wave emission, just to name a few. Nearly all major observational facilities are developed to test theory. Notable examples in physics include the LHC and LIGO, both of which confirmed theoretical predictions (the Higgs boson and gravitational waves) and generated Nobel prizes.

In many cases, extensive theory is needed just to measure a signal. For instance, the signals of merging black holes measured by LIGO would not be detected without matched filtering against theoretically generated waveform templates. At present, we are still limited in our ability to detect many types of signals (e.g., eccentric mergers) due to the lack of available templates and a lack of theorists, funding, and computational resources to generate them. If our community continues to invest so heavily into such facilities, shouldn't we also be funding the theoretical research necessary to extract the most science from our investments?

\subsection{Interpretation of Observations}

Theory also is vital to learn from observations. Consider the CMB. This relatively simple data set (a measurement of temperature across the sky) has fueled much of the field of cosmology. But none of our current understanding (e.g., the big bang, cosmic nucleosynthesis, the $\Lambda$-CDM paradigm) would exist without extensive theoretical efforts to interpret CMB data.

Most fields in astrophysics have not received as much theoretical attention, and enormous advancement could be achieved (with data already obtained) if more support for theoretical analysis was available. This need will only grow in the coming decades with the explosion of unprecedentedly large datasets that will require the trained eye of theorists to find order and physical insight.

We note though that we are not advocating for a horde of theorists that are tasked to simply ``match" the data from upcoming missions.  Rather, the goal should be a deeper understanding. A successful investment in theory will reinvigorate the field, particularly in areas where observations are at present too rudimentary to test predictions.

\section{Strategic Plan}

The strategy for the future needs to be as bold and visionary in the theoretical domain as it is in the observational and instrument domains.  One can cite many incredible machines/missions for the future:  JWST, TMT, GMT, LSST, Lynx, HABEX, OST, LUVOIR, WFIRST, CMB-S4. The list goes on.  Missing from this long catalog is any explicit support for theoretical astrophysics.  Indeed, over this same period of tremendous growth, support for theory has declined precipitously.  Even the highly successful NASA Astrophysics Theory Program has seen dramatic cuts, just as it was created.  To deny the need for this is to pretend that progress comes from experiment in isolation.  The consequence of such thinking will be the intellectual death of our field.

\subsection{National Theory Program} 
The only nation in the world with a National program in theoretical astrophysics is Canada, the Canadian Institute for Theoretical Astrophysics (CITA).  CITA has long been regarded as an intellectual jewel of Canada and we should take lessons from this successful program from our Northern neighbor.  The US investment in theoretical astrophysics should be proportionate to the US investment in national observational facilities and should thus be expanded greatly and reflect our national commitment to astrophysics research.  The national program should have a central hub, but should serve all theory groups in the nation.  As part of this, we should use the National Laboratory infrastructure to continue and expand support for theoretical astrophysics research at these sites and at their Leadership Computational Facilities to serve U.S. national science interests and capabilities.

\subsubsection{NAS Fellows in Theoretical Astrophysics}
The establishment of a national fellowship in theoretical astrophysics is a priority for 2020-2030.  In contrast to programs like the extremely successful Hubble Fellowship program (or Chandra and Sagan fellowships) that are tied to a specific mission or a specific domain area, the national fellowship in theoretical astrophysics should be open to all theorists regardless of whether their theoretical program makes direct contact with a specific observational initiative or not.  It is indeed critical to foster theory both within {\it and} outside of the auspices of a specific observational mission or program. Support of theory not limited to current missions could lead to the inspiration of future missions and facilities that would not be thought of otherwise.  These fellowships should be directed at {\it both} graduate and post-graduate programs.

\subsubsection{NAS Symposia In Theoretical Astrophysics}

Theorists need to argue.  Unlike engineering, where there are multiple solutions, some more or less elegant than others but many that ``do the job", theoretical work is either correct or incorrect.  It is thus critical that theorists have a forum for vigorous argument.  Meetings such as the AAS do not provide a forum for theorists as the speaking format is only conducive to the presentation of results and not the argument about them.  

The symposia should be of general interest and should focus on several topics each year on a rotating basis.  This high-interaction format has been successfully implemented at the Kavli Institute for Theoretical Physics in California and the Aspen Institute in Colorado.  Those venues could (and should) be directly supported for the NAS Symposia, but other venues could (and should) be considered.  The essential point is that theorists, in particular, need a venue for discourse that is not overly stifled by the current conference format which is focused on presentation rather than confrontation\footnote{We note that this recommendation is highly subfield dependent}.

\subsubsection{Theory-Observation Co-Investment}

Successful investment portfolios require a combination of assets to ensure future wealth and insure against market uncertainties.  While many observational programs are extremely ambitious, we recognize that the cost and schedule of these programs are difficult to predict accurately.  With increasingly long lead times for next generation programs, it is important that the community explicitly adopt a program of co-investment in theory alongside observational missions.  The level of this co-investment may vary depending on the nature and scale of the program, but the current default (of 0\%) is unhealthy and exposes the national facilities to tremendous risk -- both in underutilized scientific potential as well as community atrophy and apathy over the long periods required to make the discovery machines of the future.

As part of this recommendation, we recommend a Summer School program for theoretical apprentices interested in making predictions for upcoming facilities.  There currently exists a very large number of summer schools for young researchers that are observation focused, especially in emerging fields, that train apprentices in data reduction and analysis techniques.  By contrast, there are almost no summer schools associated with theoretical tools with the notable exceptions of the MESA Summer School and the Kavli Summer Program in Astrophysics. Where are the opportunities for students to learn the tools of theory, in particular, in areas that require more than mathematics and physics domain knowledge (for example, numerical hydro, radiative transfer, dynamics).  We recommend that funding be allocated for summer schools for theorists to support these flexible and targeted training grounds.

\bigskip
\noindent
\begin{tcolorbox}[colback=blue!5!white,colframe=blue!75!black,title=NAS ACTION]
    \begin{itemize}[leftmargin=0.4cm]
    \setlength\itemsep{0.1em}
    \item {The National Academy should form, or recommend the formation of, a National Program in Theoretical Astrophysics with multiple nodes and representatives across the nation.}
    \item{The National Academy should establish, or recommend the establishment of, Theory Fellowships, distinct from Hubble, Einstein, or other fellowship programs, that are exclusively for theoreticians at graduate and postdoctoral stages.}
    \item{The National Academy should establish or recommend the establishment of Theory Symposia}
    \item{The National Academy should explicitly recommend a theory component for all large facilities recommended as part of Astro 2020}
    \end{itemize}
\end{tcolorbox}

\subsection{Infrastructure for Theory}

There are several ways in which Theory infrastructure must be supported.  The first, of course, is personnel.  While NAS cannot recommend the wholesale manufacture of FTE lines at universities, it can support infrastructure at the local and national levels to leverage the academic workforce.

\subsubsection{Local Infrastructure}
As national facilities have grown to exa-scale, they effectively require local mid-scale computing infrastructure to enable their use. Mid-scale facilities are needed for code development, testing, scaling tests, and a huge range of analysis projects (e.g. radiation transfer, structure finding, etc.) which are too expensive to run on a single workstation but cannot be deployed in ``production" calculations without prior work. Without local facilities, it is difficult if not impossible to demonstrate the readiness level of code development, scaling, and testing required for large allocations on national facilities.   National facilities on larger scales are also increasingly unable to accommodate small testing protocols (e.g., analysis). Moreover, large national facilities typically guarantee data storage for only the duration of the main proposal -- any usage of the data generated beyond the grant period requires local mid-scale facilities. 

These resources are generally funded in one of two ways: either (1) groups directly pay for equipment (purchase of a mid-scale compute cluster) and support (paying administration costs as salaries or fees, which are usually comparable in cost to the equipment buy over a few-year equipment lifetime), or (2) increasingly, as universities centralize compute support to leverage economies of scale and cloud computing becomes more widely useful, groups pay a bundled cost to the university or external supplier for computing (either paying 'up front' via a subscription or buy-in fee, or paying as-you-go per compute hour). However the current grants infrastructure, especially for small (e.g., single-PI) theory grants and super-computing grants, provides no support for either of these infrastructures. Option (2) in particular, which is rapidly becoming the norm in both business and academic systems, does not even fit into the existing budget categories of most grant agencies. It is imperative that grants supporting the huge time and effort, for both scientific analysis and supercomputing time, provide a category to financially support the required local infrastructure expenses, especially the payment of administrative or per-hour compute costs.

\subsubsection{Legacy Support for Theory Products}  
Theory grants and national super-computing centers spend millions of dollars and tremendous effort and resources supporting the generation of theoretical models and predictions. A major scientific goal of these is, without exception, providing predictions or comparisons to current and future observational data. But there is no support for the archiving or access of such data. With super-computing proposals, data management plans are often required, but there is no infrastructure provided by any of the funding agencies for making that data available in useful form after the grant period expires. There is also no way to ensure theoretical data products are made public, in the way that is now viewed as an obvious (and tremendously important) requirement for observational products. Given that there is already a large infrastructure in place to support observational data sets, it is a natural and straightforward step to support mock theoretical data products of said data. This requires no new fundamental infrastructure, simply support for additional data sets, the creation of which is already being funded -- essentially, simply providing support for access of said products (after financial and computing resources are awarded for their creation) in the same way as is already done for their observational equivalents. This would also tremendously increase the impact of theoretical products, by making them not just more readily accessible to observers, but accessible via the same interactive data archival systems which they already use to access the equivalent observational products, catalogues, images, spectra, etc.

\bigskip
\noindent

\begin{tcolorbox}[colback=blue!5!white,colframe=blue!75!black,title=NAS ACTION]
\begin{itemize}[leftmargin=0.4cm]
\setlength\itemsep{0.1em}
    \item {Grant support for local infrastructure and maintenance should be recommended as a priority for 2020-2030 within the AST and NASA grant programs directly.}
    \item{Explicit competitions for replacement infrastructure for astrophysics theory should be recommended and regularly conducted along with mid-scale infrastructure calls.}
    \end{itemize}
\end{tcolorbox}

\subsection{Program Support}
Both NSF and NASA have the resources to direct toward this vision if they have the community support to do so.  Investment in theory should be made an explicit priority and should be done in a proportion $\alpha$ to the investment in the observatories of the future.  We do not speculate what $\alpha$ should be in this document.  We suggest NAS undertake a study to determine this number, but we argue it should be no less than $10\%$.

\section{Conclusions}
The national investment in observational infrastructure is growing without bound, but in sharp contrast, the national investment in theory has remained constant for decades.  This situation threatens the health and vitality of our field. The goal of astronomy and astrophysics is not just to catalog the universe, but to understand how it works.  Without broad support of theoretical work, ranging from theoretical predictions that suggest new observations or instruments, to theoretical interpretation and modelling of existing data, astronomy will be reduced to an expensive and aimless cosmic census.

\newpage
\end{document}